\documentclass{ifacconf}

\usepackage{eufrak} 
\usepackage{graphicx}      
\usepackage{natbib}        
\usepackage{amsmath}
\usepackage{amsfonts}

\newcommand{\parderiv}[2]{{ \frac{\partial #1}{\partial #2} }}
\newcommand{\deriv}[2]{{ \frac{\textnormal{d} #1}{\textnormal{d} #2} }}
\newcommand{\euler}{\textnormal{e}}
\begin{document}
\begin{frontmatter}

\title{Nonlinear feedback stabilisation and stochastic disturbance suppression of actively Q-switched lasers\thanksref{footnoteinfo}} 

\thanks[footnoteinfo]{The project 'E! 114831 ESPRESSO' is supported by the Eurostars-2 programme and has received funding from the European Community and the Austrian Federal Ministry for Digital and Economic Affairs (FFG grant agreement No. 884176).}

\author[First]{Lukas Tarra} 
\author[First]{Andreas Deutschmann-Olek} 
\author[First,Second]{Andreas Kugi}

\address[First]{Automation and Control Institute, TU Wien, Gu\ss hausstra\ss e 27-29, 1040 Vienna, Austria}
\address[Second]{Center for Vision, Automation \& Control, Austrian Institute of Technology, Giefinggasse 4, 1210 Vienna, Austria}

\begin{abstract}                
Actively Q-switched lasers are widely used tools which are required to produce stable output pulse energies for many applications. In this paper, a model-based control concept for actively Q-switched lasers is presented  which stabilises their nonlinear pulse-to-pulse dynamics and rejects stochastic disturbances arising from amplified spontaneous emission. The feasibility of the control task is demonstrated to strongly depend on the design of the semi-active prelasing approach. In contrast to state-of-the-art hardware-based controllers, the proposed concept is flexible and cost-effective as it is not tailored to specific operation parameters. 
\end{abstract}

\begin{keyword}
Q-switched laser, Feedback stabilization, Control of bifurcation and chaos, Disturbance rejection, Estimation and filtering
\end{keyword}

\end{frontmatter}

\section{Introduction}
\label{section:introduction}
Pulsed lasers with output pulse widths in the nanosecond range are important tools for material processing tasks such as metal cutting \citep{q_switch_application}, medical skin treatment \citep{medical_application}, and imaging applications such as LIDAR \citep{lidar_application}. An established technique for achieving such pulses is active Q-switching, where an active medium (for example an Nd:YAG crystal that is continuously pumped, e.g., by a laser diode) is placed within an optical cavity of externally controllable quality. Each laser pulse is the result of one switching cycle: at first, the cavity is kept at low quality such that the atomic excitation of the active medium increases due to optical pumping. After the cavity is switched to high quality, spontaneous emission and subsequent induced emission lead to a rapid build-up of optical energy within the cavity. At the end of the switching cycle, the cavity is opened again, and the optical energy is released as a laser pulse \citep{koechner_book}. 
Laser engineers are currently pursuing the goal to push the pulse repetition rate, i.e. the rate at which switching cycles are performed, to values around 1 MHz. High repetition rates increase the processing speed in a variety of applications, e.g., material removal can be very finely controlled while a reasonable processing speed can be maintained \citep{desire_for_high_rep_application}.    
In this regime, users are facing two major unwanted influences on output pulse energies which compromise the system's performance and can even damage the laser itself. 
On the one hand, the atomic excitation population of the active medium couples subsequent laser pulses at high repetition rates. The resulting nonlinear pulse-to-pulse dynamics often exhibit nontrivial limit cycles due to period-doubling bifurcations and even deterministic chaos \citep{instability_experiment}. On the other hand, amplified spontaneous emission, which builds the basis of the energy build-up, is a stochastic process which introduces stochastic fluctuations to an output pulse's shape and its total energy \citep{stochastic_experiment}.
Such energy fluctuations in particular limit sensing applications which are sensitive to the optical intensity. Hardware-based attempts have been made \citep{fluorescence_feedback} to mitigate those dynamic and stochastic influences, but such concepts are tailored to specific operation parameters and suffer from little flexibility and high costs. \\
In this paper, we present a model-based nonlinear feedback approach which ensures global asymptotic stability of the closed-loop system in the deterministic case and rejects the influence of stochastic disturbances which are observed in real time. In Sec. \ref{section:mathematical_model}, we summarize a mathematical model of actively Q-switched lasers which was recently presented in \citep{Tarra_model}. Possible actuators and measurable quantities are discussed in Sec. \ref{section:controller_and_estimator_design}, where we then present the controller and disturbance estimator design. The performance of the proposed approach is assessed in a simulation study in Sec. \ref{section:simulation_results}. Finally, an outlook on further developments of the application is given and conclusions are drawn in Sec. \ref{section:conclusion_and_outlook}.
\section{Mathematical model}
\label{section:mathematical_model}
\begin{figure}
\begin{center}
\includegraphics[width=0.7\linewidth]{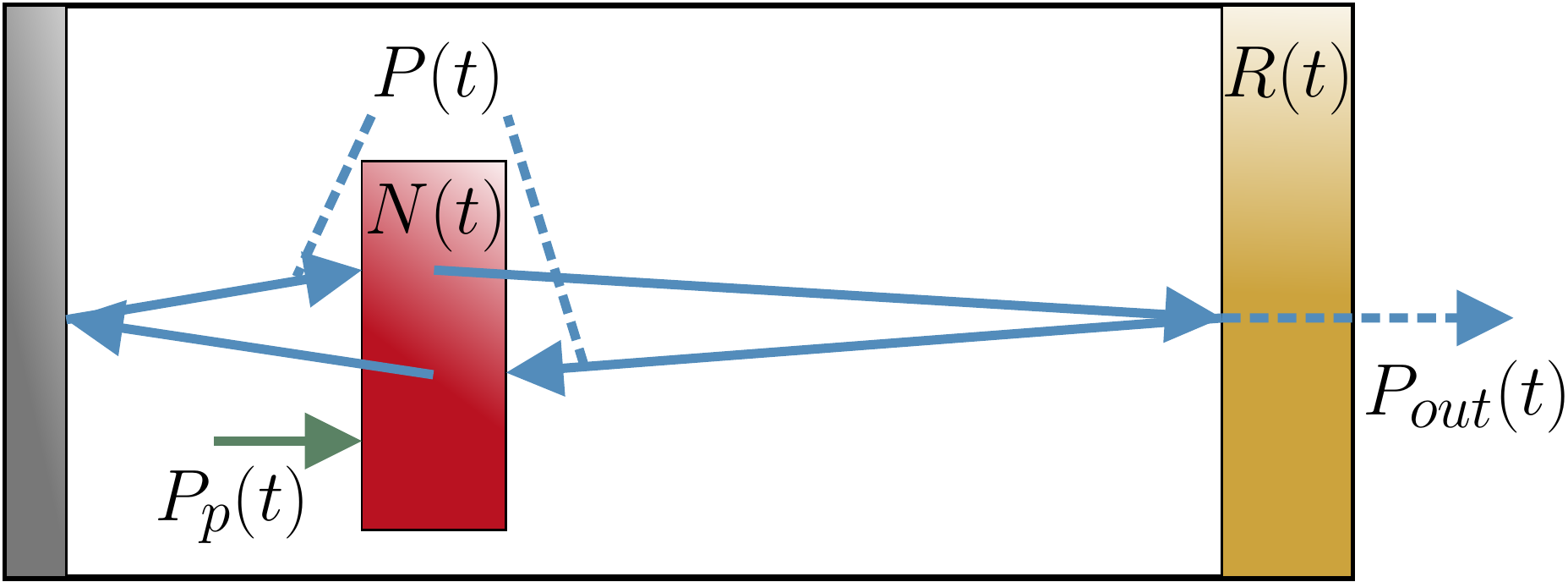} 
\caption{Simplified sketch of a Q-switched laser cavity. Pump and laser beams are represented by green and blue arrows, respectively. The red box stands for the active medium and the yellow box for the Q-switch.} 
\label{fig:cavity_skizze}
\end{center}
\end{figure}

The fundamental principle of Q-switching was already explained in Sec. \ref{section:introduction}. In Fig. \ref{fig:cavity_skizze}, one can see a simplified sketch of the laser cavity and the active medium with the pump power $P_p(t)$, the part $P(t)$ of the intracavity power which is directly applied to the active medium, the excitation population $N(t)$, and the output power $P_{out}(t)$. The reflection coefficient $R(t) \in (0,1)$ can be switched between different values to change the Q-factor of the cavity. The higher $R(t)$, the more radiation remains within the cavity. 

\subsection{Effective single-mode CTNM}
\label{section:effective_single_mode_CTNM}
In \citep{Tarra_model}, a stochastic, spatially lumped model - called the continuous-time nonlinear model \linebreak(CTNM) therein - was derived, starting from a general, first-principles-based, deterministic, distributed-parameter description widely used in the literature, e.g., in \citep{stateofart_2021}. One important feature of this model is that a number of independent longitudinal lasing modes $j \in \{1, \dots, M\}$ is started by one compound Poisson process $\mathfrak{C}_j(t)$ each representing stochastic fluctuations due to amplified spontaneous emission. The resulting randomly time-varying spectral power profile leads to $M+1$ dynamically evolving quantities, namely $M$ lasing modes competing for amplification which originates from one excitation population $N(t)$. This model was compared to spatially distributed models, models which include several atomic excitation densities, and first measurements on an experimental system in \citep{Tarra_model}. Good agreement was demonstrated in all cases.
If one is not interested in the exact spectral distribution of energy within the cavity or investigates spectrally selective single-mode lasers, the CTNM can be replaced by an effective single-mode model which reads as \citep{Tarra_model}
\begin{subequations}\label{equ:CTNM_avg}
\begin{align}
\deriv{ N(t)}{t} &=  \frac{b \lambda_{p}}{h c A_p} \bigg( 1- \euler^{ \sigma_{p} \left(a N(t) -L N_{dop} \right) -\alpha_pL } -\alpha_pL \bigg) P_p(t) \notag \\
& - b \gamma N(t) - \frac{b}{h c A_s}  \bigg( \sum_{k=0}^\infty q_k (N(t))^k -\Lambda \bigg) P(t)\label{equ:CTNM_avg_N} \\
\begin{split}  \deriv{P(t)}{t} &= \frac{2q_1}{q_0 t_{RT}} N(t) P(t) -  \left( \frac{1}{\tau}-\frac{\ln(R(t))}{t_{RT}} \right) P(t)   \\ 
 &+  \frac{ 2}{t_{RT}} \Big(  \frac{\Delta\Omega}{4\pi} \Sigma(t) +\alpha_{RS}L P(t) \Big) \ ,\end{split} \label{equ:CTNM_avg_P}
\end{align}
\end{subequations}
where the output power is in good approximation given by
\begin{align}\label{equ:P_out_CTNM}
P_{out}(t) =  \frac{\left( 1-R(t) \right)}{R(t)} P(t) \frac{\euler^{\sigma N(t)-\alpha L } }{\frac{1}{\sqrt{\eta}R(t)} + \euler^{ \sigma N(t)-\alpha L } }   \ . 
\end{align}
The parameter $\lambda_p$ is the wavelength of the pump beam, $h$ and $c$ are Planck's constant and the vacuum speed of light, $A_p$ and $A_s$ the cross-sectional areas of the pump and laser beams, $\sigma_p$ and $\sigma$ the scattering cross-sections of the pump and radiation transitions, $\alpha_p$ and $\alpha$ are loss coefficients of the active medium, $\gamma$ the population relaxation rate, $N_{dop}$ the atomic dopand density, $L$ the length of the active medium along the direction of beam propagation, $t_{RT}$ the cavity round-trip time for light, $\alpha_{RS}$ the Rayleigh backscattering coefficient, $\eta$ the cavity loss coefficient, $\Delta \Omega$ the solid angle at which spontaneously emitted photons are captured by the beam and $a$ and $b$ are thermalisation constants stemming from singular perturbations. In (\ref{equ:CTNM_avg}), the effective parameters $q_k$, $\Lambda$ and $1/\tau$ stand for gain/depletion characteristics, the average lasing wavelength and effective static power losses, respectively. The stochastic noise $\Sigma(t)$ combines effective spontaneous emissions arising from all noise channels.
The effective single-mode model represents a computationally efficient model which describes the overall evolution of energy within an actively Q-switched laser. Due to thermal effects in real lasers, its parameters must be identified from measurement data, as will be further discussed in Sec. \ref{section:conclusion_and_outlook}.

\subsection{Pulse-to-pulse dynamics}
\label{section:pulse-to-pulse_dynamics}

\begin{figure}
\begin{center}
\includegraphics[width=0.88\linewidth]{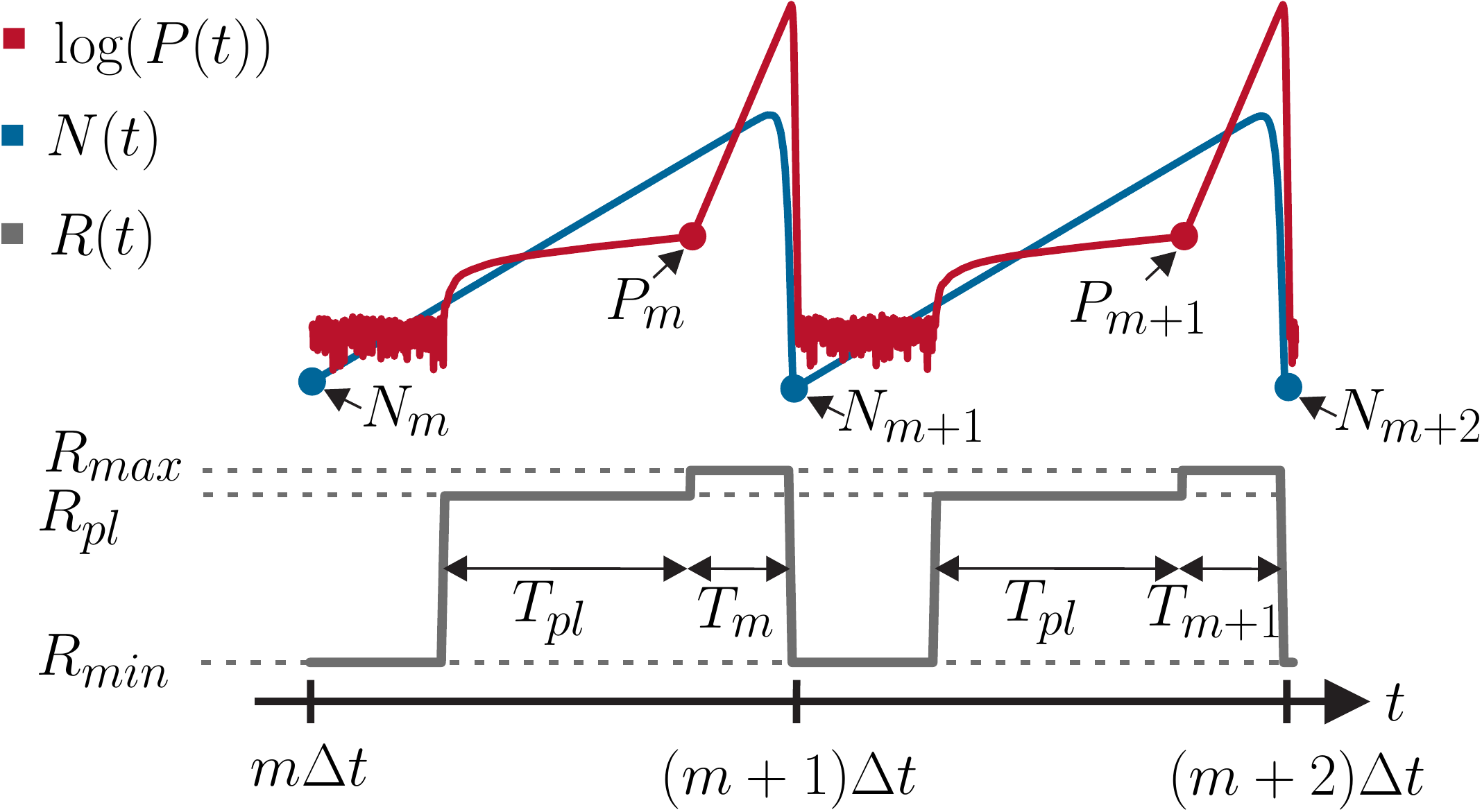} 
\caption{Illustration of the continuous dynamics of (\ref{equ:CTNM_avg}) leading to the pulse-to-pulse dynamics (\ref{equ:pulse-to-pulse-dynamics}). The intracivity power (red) exhibits separate growth rates in the prelasing phase (duration $T_{pl}$) and the high-Q phase (duration $T_m$). Populations $N_m$ and $N_{m+1}$ are coupled via subsequent growth and depletion (blue).} 
\label{fig:cycle_graphic}
\end{center}
\end{figure}

The real-time behaviour of actively Q-switched lasers can be mathematically described by (\ref{equ:CTNM_avg}). As mentioned in Sec. \ref{section:introduction}, the pulse-generating cyclic operation of the system results from the periodic switching of $R(t)$, which is why the switching frequency $f_{switch}$ is equal to the pulse repetition rate. Therefore, the time increment between two subsequent cycles is given by 
\begin{align}\label{equ:Delta_t_definition}
\Delta t := 1/f_{switch} \ .
\end{align}
In standard Q-switching, $R(t)$ is switched between a low Q-value $R_{min}$ and a high Q-value $R_{max}$. The corresponding time intervals are called low-Q phase (or pump phase) and high-Q phase (or duty cycle). Pulses obtained from standard Q-switching are subject to substantial stochastic fluctuations since no intracavity power can build up during the low-Q phase, and consequently, the seeding power for the high-Q phase is given entirely by the current value of the stochastic noise $\Sigma(t)$ in (\ref{equ:CTNM_avg_P}) and hence is fundamentally random. To combat this issue, one can employ prelasing, which is characterised by a phase of intermediate cavity quality $R_{pl}$ and a duration $T_{pl}$, see the lower part of Fig. \ref{fig:cycle_graphic}, where some amplification of the intracavity power can occur \citep{koechner_book}. Such a slower build-up (compare the slopes and thus the exponential growth rates of $P(t)$ in Fig. \ref{fig:cycle_graphic}) has the effect that stochastic fluctuations leading to the high-Q phase can be averaged out to some degree. However, this effect is limited, especially at high repetition rates where $T_{pl}$ is necessarily short. In order to still be able to compensate for stochastic fluctuations from spontaneous emission, one could continuously adapt $R_{pl}$ during the prelasing phase, yielding a well-defined and constant intracavity power value at the beginning of the high-Q phase.  Such active prelasing requires additional real-time-adjustable optical elements within the cavity, which is usually undesired due to additional costs and inevitable static losses. \\
As an alternative, we propose semi-active prelasing, where the timing of the Q-switch transition from prelasing to the high-Q phase ($R_{pl} \rightarrow R_{max}$) is adjusted based on information from continuous monitoring of the slower energy build-up. That way, stochastic fluctuations are taken into account by real-time measurement while only a shift in the timing of the high-speed circuitry needs to be controlled, which requires no additional optical elements within the cavity. Although difficult at such high repetition rates, one could finally use the pump power $P_p(t)$ as a control input, but this option will not be discussed further in this work. Both active and semi-active prelasing rely on the fact that $P(t)$ and $P_{out}(t)$ can be measured as small fractions of their total values which are transmitted through optical mirrors within and outside the cavity, respectively. Depending on space within the cavity and whether the active medium is cooled, $N(t)$ can also be directly determined by fluorescence measurement or through transmission of a weak probe beam \citep{Vinzenz}.  
While the laser's behaviour during a switching cycle is a continuous-time dynamical system described by (\ref{equ:CTNM_avg}), the pulse-to-pulse dynamics couple subsequent pulses and thus occur in discrete time, with a sampling time given by (\ref{equ:Delta_t_definition}) (see Fig. \ref{fig:cycle_graphic} for an illustration of this connection). As state variable for the pulse-to-pulse dynamics, we use the cycle's initial population given by
\begin{align*}
N_m := N(m\Delta t) \ . 
\end{align*}
The resulting pulse energy of the $m$-th cycle is defined as
\begin{align} \label{equ:E_out}
E_{m} :=  \int_{(m-1)\Delta t}^{m\Delta t} P_{out}(t) \ \textnormal{d}t \ ,
\end{align}
where $P_{out}(t)$ is given by (\ref{equ:P_out_CTNM}). The coupling between two pulses through the populations $N_m$ and $N_{m+1}$ can be understood as follows: a large $N_m$ leads to a strong $m$-th pulse (large value of $E_m$ in (\ref{equ:E_out})) which depletes a substantial amount of population through the last term in (\ref{equ:CTNM_avg_N}) such that $N_{m+1}$ and therefore $E_{m+1}$ are small etc.
As discussed above, semi-active prelasing employs the duration of the $m$-th high-Q phase as input, which shall be denoted by $T_m$. The intracavity power $P_m$ at the beginning of the $m$-th high-Q phase is related to $P(t)$ by
\begin{align*}
P_m := P\left((m+1)\Delta t - T_m \right) = P_m (N_m,T_m) \ .
\end{align*}
$P_m$ is a random variable since it is obtained from stochastic evolution of (\ref{equ:CTNM_avg}). Further, it depends on $N_m$ and $T_m$ because the value of $N_m$ determines the eventual gain and a longer high-Q phase $T_m$ leads to a shorter prelasing phase and thus less amplification time for $P_m$ (see Fig. \ref{fig:cycle_graphic}). Once the value of $P_m$ is set and the high-Q phase begins, the rest of the switching cycle can be regarded as deterministic, i.e. as an evolution of (\ref{equ:CTNM_avg}) with $\Sigma(t)$ replaced by 
\begin{align}\label{equ:stochastic_process_mean}
\mu(t) := \mathbb{E}(\Sigma(t)) \ ,
\end{align}
as stochastic variations due to spontaneous emission are then negligible compared to the sharply rising amplification curve of $P(t)$. \\
Mathematically, the pulse-to-pulse dynamics can therefore be stated as the discrete-time dynamics
\begin{subequations} \label{equ:pulse-to-pulse-dynamics}
\begin{align}
N_{m+1} &= f(N_m,P_m(N_m,T_m),T_m) \label{equ:pulse_to_pulse_dynamics_f} \\
E_m &= h(N_m,P_m(N_m,T_m),T_m) \ . \label{equ:pulse_to_pulse_dynamics_h}
\end{align}
\end{subequations}                                                                                                                                                                                                                                                                                                                                                                                                                                                                                                                                                                                                                                                                                                                                                                                                                                                                                                                                                                                                                                                                                                                                                                                                                                                                                                                                                                                                                                                                                                                                                                                                                                                                                                                                                                                                                                                                                                                                                                                                                                                                                                                                                                                                                                                                                                                                                                                                                                                                                                                                                                                                                                                                                                                                                                                                                                                                                                                                                                                                                                                                                                                                                                                                                                                                                                                                                                                                                                                                     
The concrete functional form of (\ref{equ:pulse-to-pulse-dynamics}) may be either obtained from approximate analytic solution of (\ref{equ:CTNM_avg}), data-based methods or numerical forward-simulation of (\ref{equ:CTNM_avg}). In this paper, we will follow the latter approach. We interpret the random variable $P_m(N_m,T_m)$ as a stochastic disturbance as it is determined by the current realisation of the noise term $\Sigma(t)$ in (\ref{equ:CTNM_avg_P}) and triggers the subsequent pulse amplification.

\section{Controller and estimator design}
\label{section:controller_and_estimator_design}

\subsection{Control task}
\label{section:Control_task}
As mentioned in Sec. \ref{section:introduction}, pulse energies are affected by both dynamic pulse-to-pulse instabilities of the active medium and stochastic disturbances due to amplified spontaneous emission. In terms of the dynamical system (\ref{equ:pulse-to-pulse-dynamics}), this means that on the one hand, the deterministic population dynamics
\begin{align}\label{equ:deterministic_pulse_to_pulse}
N_{m+1} = f(N_m,P_s(N_m,T_m),T_m) 
\end{align}
with
\begin{align}
P_s(N_m,T_m) := \mathbb{E}\left(P_m(N_m,T_m)\right) \label{equ:P_s_definition}
\end{align}
exhibit an unstable steady state $N_s$ defined by
\begin{align*}
N_{s} &= f(N_s,P_s(N_s,T_s),T_s) 
\end{align*}
for a given nominal value $T_s$. On the other hand, $P_m$ varies stochastically such that both (\ref{equ:pulse_to_pulse_dynamics_f}) and (\ref{equ:pulse_to_pulse_dynamics_h}) are affected. Therefore, the resulting control task can be summarized as follows: use $T_m$ to stabilise (\ref{equ:pulse_to_pulse_dynamics_f}) around the steady-state population $N_s$ and mitigate the effect of stochastic fluctuations $P_m - P_s$ using measurements of the intracavity power $P(t)$. 


\subsection{Nonlinear GAS controller}
\label{section:nonlinear_gas_controller}

To globally asymptotically stabilise (hence GAS) the pulse-to-pulse dynamics (\ref{equ:pulse_to_pulse_dynamics_f}), we will follow the approach used in \citep{deutschmann_gas_feedback} for regenerative amplifiers. For now, we will focus on the stabilisation of (\ref{equ:deterministic_pulse_to_pulse}) and disregard stochastic influences as their mitigation will be dealt with in Sec. \ref{section:linear_disturbance_feedback}. 
If we use the nonlinear feedback law $T_m = g(N_m)$, a sufficient condition for global asymptotic stability of the closed-loop dynamics
\begin{align*}
N_{m+1} = f_{CL}(N_m) =: f(N_m,P_s(N_m,g(N_m)),g(N_m)) 
\end{align*}
is that for every possible $N_m$, we have 
\begin{align} \label{equ:stability_condition}
\left| \deriv{f_{CL}(N_m)}{N_m} \right| < 1 \ .
\end{align}
This can be shown via Lyapunov's direct method \citep{discrete_time_systems}, e.g., with the Lyapunov function $V(N_m) := |N_m-N_s|$. By taking the total derivative of $f_{CL}$ w.r.t. $N_m$ and setting the nominal steady-state input $T_s$ as an initial condition, one directly finds a differential equation for $g(N_m)$ which reads as
\begin{subequations} \label{equ:gas_controller_ode}
\begin{align}
 \deriv{g}{N_m} =& \frac{ \deriv{f_{CL,des}}{N_m} - \parderiv{f}{N_m } - \parderiv{f}{P_m} \parderiv{P_s}{N_m} }{   \parderiv{f}{T_m} + \parderiv{f}{P_m} \parderiv{P_s}{T_m} } \\
 g(N_s) =& \  T_s \ ,
\end{align} 
\end{subequations}
with the desired closed-loop behaviour $f_{CL,des}(N_m)$. For a control law which is minimally invasive to the system's natural dynamics while still ensuring (\ref{equ:stability_condition}) by flattening the falling slope of the uncontrolled system $f(N_m,P_s(N_m,T_s),T_s)$ (c.f. Fig. \ref{fig:controller_design}), we make the choice
\begin{align}\label{equ:desired_closed_loop_slope}
 \deriv{f_{CL,des}}{N_m} = \Bigg\{ \begin{array}{c}
 -1+\alpha \  , \hspace{0.7 cm} \  \parderiv{f}{N_m} + \parderiv{f}{P_m} \parderiv{P_s}{N_m}  < -1+\alpha \\
\hspace{-2.4 cm} \parderiv{f}{N_m} +\parderiv{f}{P_m} \parderiv{P_s}{N_m}  \hspace{0.7 cm}  \textnormal{else} \ ,
 \end{array}
\end{align}
with the stability margin $ 0< \alpha < 1$. Such a minimally invasive approach is chosen for the sake of robustness. From solving (\ref{equ:gas_controller_ode}), we obtain a nonlinear function $g(N_m)$ which can be stored as a lookup table, enabling fast online implementation.

\subsection{Linear disturbance compensation}
\label{section:linear_disturbance_feedback}

Assuming that we have an estimate $\hat{P}_m$ of the current intracavity power $P_m$ at the switching time $(m+1)\Delta t - T_m$ (see Fig. \ref{fig:Prelasing_Kalman_illustration}), we can reject its stochastic influences in a dead-beat sense. Since the deviation of $P_m$ from its expectation value $P_s(N_m,T_m)$ is assumed to be small, a linear correction is deemed sufficient. Augmenting the nonlinear feedback law $g(N_m)$ by the linear ansatz 
\begin{align}\label{equ:full_control_law}
T_m = g(N_m) + k(N_m) \left( \hat{P}_m -P_s(N_m,T_m) \right) \ ,
\end{align}  
we can demand that the stochastically disturbed system with this additional compensation term in the input behaves like the stabilised deterministic system from Sec. \ref{section:nonlinear_gas_controller}. If we assume an ideal estimate $\hat{P}_m = P_m$, this reads as
\begin{align}\label{equ:demanded_P_dead_beat}
\begin{split} f \left( N_m,P_m, g(N_m) + k(N_m) \left( P_m -P_s(N_m,g(N_m)) \right) \right) \\ \stackrel{!}{=} f \left( N_m,P_s(N_m,g(N_m)), g(N_m) \right) \ . \end{split}
\end{align}
Differentiation of (\ref{equ:demanded_P_dead_beat}) w.r.t. $P_m$, application of the chain rule, and evaluation at $P_m = P_s(N_m,g(N_m))$ readily yield the linear correction factor
\begin{align}\label{equ:k_compensation_factor}
k(N_m) = -  \frac{  \parderiv{f}{P_m}(N_m,P_s(N_m,g(N_m)),g(N_m)) }{  \parderiv{f}{T_m} (N_m,P_s(N_m,g(N_m)),g(N_m)) } \ .
\end{align}

\subsection{Real-time disturbance estimator}
\label{section:Real-time_disturbance_filter}

To finally obtain a suitable estimate $\hat{P}_m$, one can utilize measurements of $P(t)$. However, one cannot simply measure up to time $(m+1)\Delta t -T_m$ since $T_m$ is determined by (\ref{equ:full_control_law}) due to $\hat{P}_m$. Additionally, real-time measurements of $P(t)$ are typically subject to significant measurement noise. 
To untangle the dependence loop and allow for some control headroom and computation time, we introduce the point in time $\bar{t}_m$ during the prelasing phase when the decision on $T_m$ is made, i.e. when the control law (\ref{equ:full_control_law}) is evaluated. This way, it is possible to increase the current $T_m$ in online operation, which is equivalent to shifting the beginning $(m+1)\Delta t-T_m$ of the high-Q phase to earlier times towards $\bar{t}_m$. Fig. \ref{fig:Prelasing_Kalman_illustration} illustrates the relation between $\bar{t}_m$, $\hat{P}_m$, and $T_m$.
Up to $\bar{t}_m$, measurements of $P(t)$ can be used to refine the estimated value $\hat{P}(\bar{t}_m)$. Since the control law (\ref{equ:full_control_law}) requires $\hat{P}_m$, this quantity needs to be predicted based on $\hat{P}(\bar{t}_m)$. Therefore, we utilize a real-time disturbance estimator which
\begin{itemize}
\item for $t \leq \bar{t}_m $ filters the noisy measurements
\begin{align}\label{equ:estimator_measurement_equation}
y(t) = P(t) + v(t)
\end{align}
to arrive at an optimal estimate $\hat{P}(\bar{t}_m)$ (blue regime in Fig. \ref{fig:Prelasing_Kalman_illustration}) and
\item predicts the value $\hat{P}_m(T_m)$ using (\ref{equ:CTNM_avg}) and $\hat{P}(\bar{t}_m)$ as its initial condition (red regime in Fig. \ref{fig:Prelasing_Kalman_illustration}).
\end{itemize}
The measurement equation (\ref{equ:estimator_measurement_equation}) is assumed to include standard white sensor noise $v(t)$ with zero mean and variance $V$. To exploit the rich theory of optimal linear estimators, we can make the following approximations which are well justified in the prelasing phase where the estimator operates: \\
(A1) Since the intracavity power $P(t)$ is too low to deplete significant population, the last term in (\ref{equ:CTNM_avg_N}) can be neglected. \\
(A2) With non-saturated populations ($aN(t) \ll LN_{dop}$), the exponential in (\ref{equ:CTNM_avg_N}) is approximately zero. \\
Starting, e.g., at $N(0)= N_m$ and using a constant pump power $P_p$, the analytic solution of (\ref{equ:CTNM_avg_N}) can then be found as
\begin{align}\label{equ:N_prelasing_solution}
N(t) =  \frac{ \lambda_{p} P_p }{\gamma h c A_p} \left( 1-\alpha_p L \right) \left( 1- \euler^{-b\gamma t} \right) + N_m \euler^{-b\gamma t}  .
\end{align}
Plugging this result into (\ref{equ:CTNM_avg_P}), we are left with a time-variant, but linear evolution of (\ref{equ:CTNM_avg_P}). As is often done in estimator design, we replace $\Sigma(t)$ - which is a sum of compound Poisson processes which depend on time via $N(t)$ - with its expectation value (\ref{equ:stochastic_process_mean}) plus Gaussian white noise of the same variance $Q(N(t))$, i.e.
\begin{align*}
\Sigma(N(t)) \approx \mu(N(t)) + w \left( 0,Q(N(t)) \right) \ .
\end{align*}
With this approximation and the dynamics of (\ref{equ:CTNM_avg_P}) with (\ref{equ:N_prelasing_solution}) being linear in $P(t)$, we can finally apply the one-dimensional time-variant Kalman-Bucy filter \citep{filtering_theory} and obtain
\begin{subequations}\label{equ:full_Kalman_filter}
\begin{align}
&\deriv{\hat{P}(t)}{t} = A(N(t)) \hat{P}(t) + \frac{C(t)}{V} \left( y(t) -\hat{P}(t) \right) +G\mu(N(t)) \label{equ:Kalman_equation}\\
&\deriv{C(t)}{t} = 2A(N(t)) C(t) + G^2 Q(N(t)) - \frac{C(t)^2}{V} \label{equ:covariance_Kalman_equation}
\end{align}
\end{subequations}
for the estimate $\hat{P}(t)$ and its error covariance $C(t)$. As introduced earlier, $V$ is the sensor noise variance, and $A(N(t))$ and $G$ are given by
\begin{align*}
 A(N(t)) &= \frac{2q_1}{q_0 t_{RT}} N(t)  -  \left( \frac{1}{\tau}-\frac{\ln(R_{pl})}{t_{RT}} \right) + \frac{2 \alpha_{RS}L}{t_{RT}} \\ 
 G &= \frac{2}{t_{RT}} \frac{\Delta\Omega}{4\pi}
\end{align*}
since the disturbance filter is only active during the prelasing phase such that $R(t) = R_{pl}$. Even though (\ref{equ:full_Kalman_filter}) is already of a simple structure due to its scalar nature, it can be simplified further by identifying different time scales. Simulations show that within the first 50 nanoseconds of prelasing (when $T_{pl} = 500$ns), before any significant intracavity power builds up, (\ref{equ:covariance_Kalman_equation}) enters a quasi-stationary regime determined by the evolution of $N(t)$. For filtering the signal (\ref{equ:estimator_measurement_equation}), it is thus an excellent approximation to replace (\ref{equ:covariance_Kalman_equation}) by the quasi-stationary solution
\begin{align*} 
C_\infty(N(t)) = A(N(t))V + \sqrt{(A(N(t))V)^2+G^2VQ(N(t))} 
\end{align*}
which can be inserted into (\ref{equ:Kalman_equation}). This yields an estimator for the intracavity power $\hat{P}(t)$ that is particularly easy to compute. For times later than $\bar{t}_m$, i.e., when (\ref{equ:Kalman_equation}) is used as a predictor, one can simply set $C(t)=0$ since the approximations (A1) and (A2) are still valid. We thus obtain an estimate $\hat{P}_m$ depending on $\hat{P}(\bar{t}_m)$ and the actual high-Q phase $T_m$, i.e., $\hat{P}_m(\hat{P}(\bar{t}_m),T_m)$.
Due to the dependence of $\hat{P}_m(\hat{P}(\bar{t}_m),T_m)$ and $P_s(N_m,T_m)$ on $T_m$ as explained, e.g., in Fig. \ref{fig:Prelasing_Kalman_illustration}, the control law (\ref{equ:full_control_law}) is actually an implicit equation in $T_m$. However, (\ref{equ:full_control_law}) can be solved offline to obtain a final two-dimensional lookup table $T_m (N_m,\hat{P}(\bar{t}_m))$, which together with the real-time disturbance estimator (\ref{equ:full_Kalman_filter}) completes the control concept presented in this work.

\section{Simulation results}
\label{section:simulation_results}

\begin{figure}
\begin{center}
\includegraphics[width=0.74\linewidth]{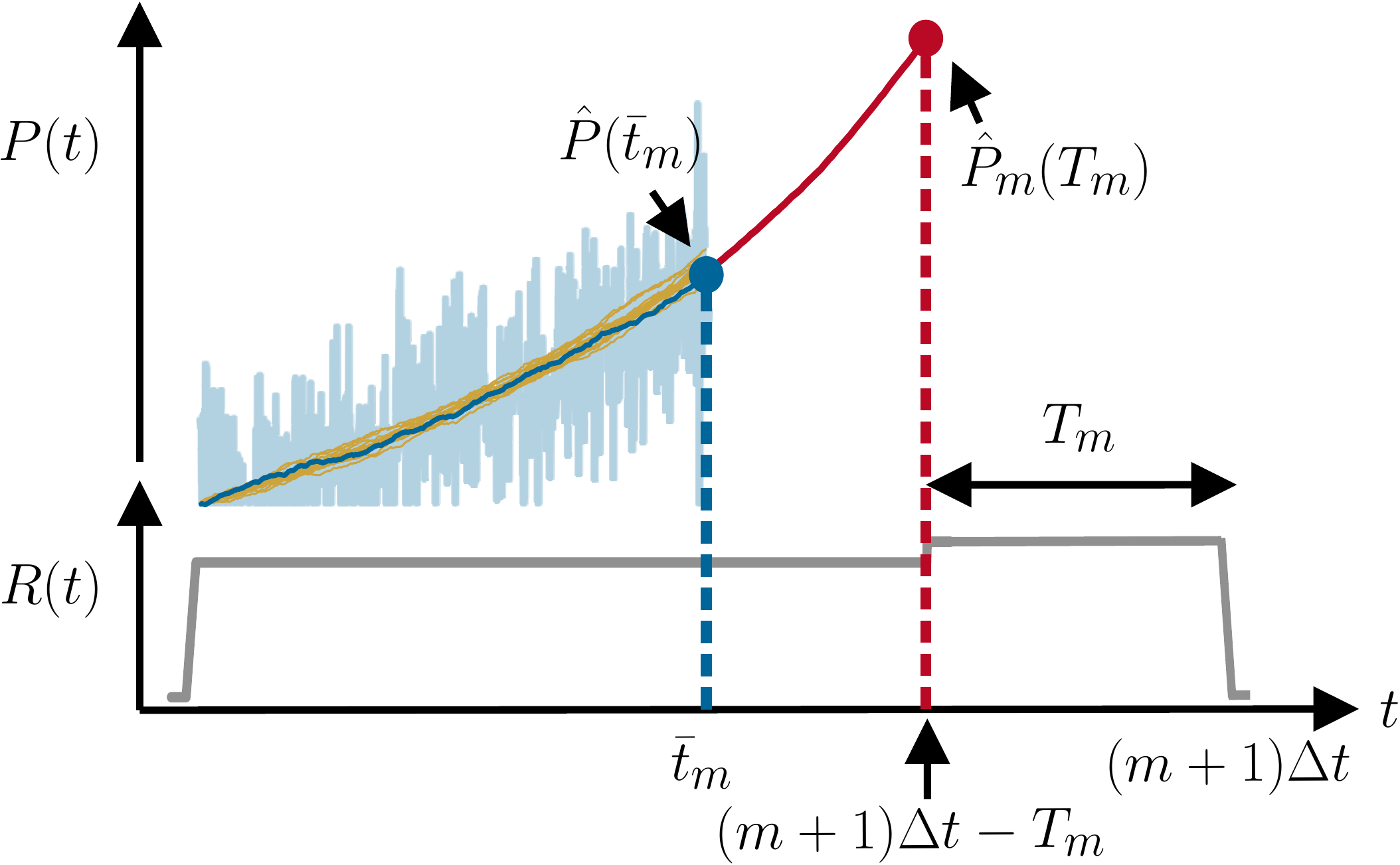} 
\caption{Illustration and simulation of the real-time disturbance estimator. Up to $\bar{t}_m$, measured data (light blue) are used to estimate $\hat{P}(\bar{t}_m)$ (blue dot). Using model knowledge (red), the disturbance $P_m$ is predicted.} 
\label{fig:Prelasing_Kalman_illustration}
\end{center}
\end{figure}

\begin{figure}
\begin{center}
\includegraphics[width=0.66\linewidth]{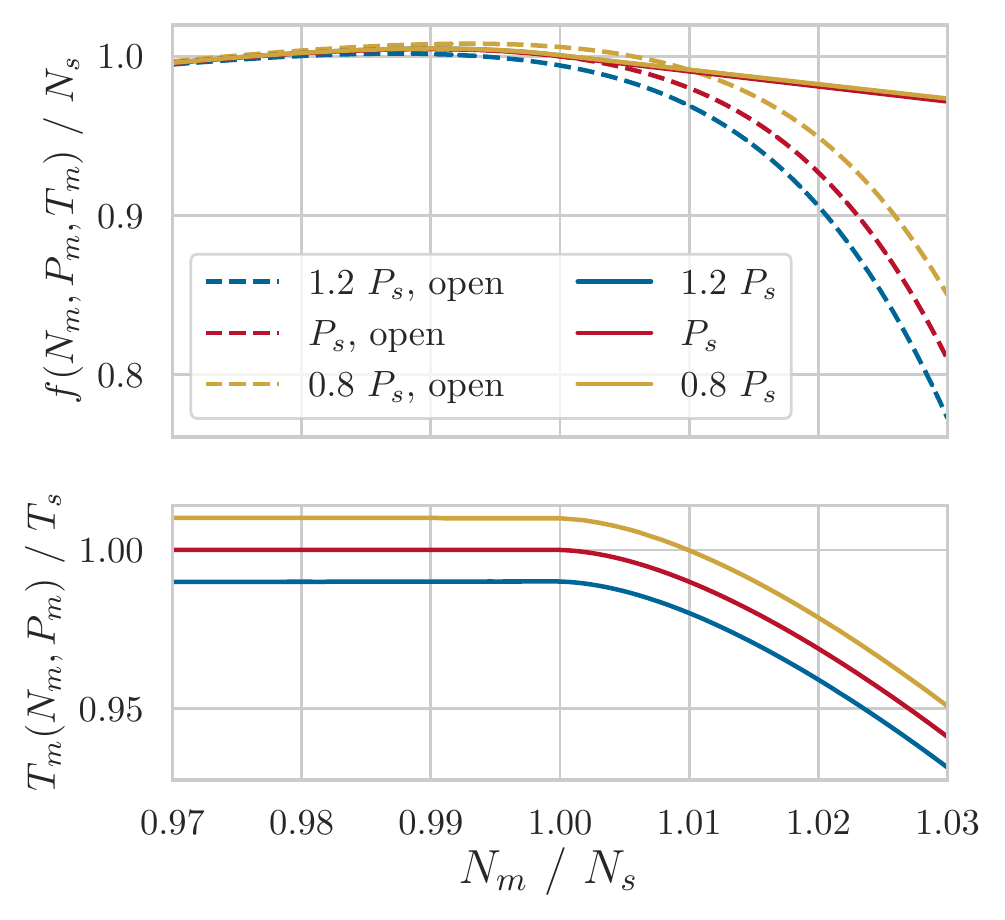} 
\caption{Illustration of the combined effects of nonlinear GAS feedback control law and linear disturbance compensation. Top: open- (dashed) and closed-loop (solid) population dynamics maps for disturbances $P_m$ above, equal to and below $P_s$. Bottom: the corresponding normalised control inputs.} 
\label{fig:controller_design}
\end{center}
\end{figure}

\begin{figure}
\begin{center}
\includegraphics[width=1\linewidth]{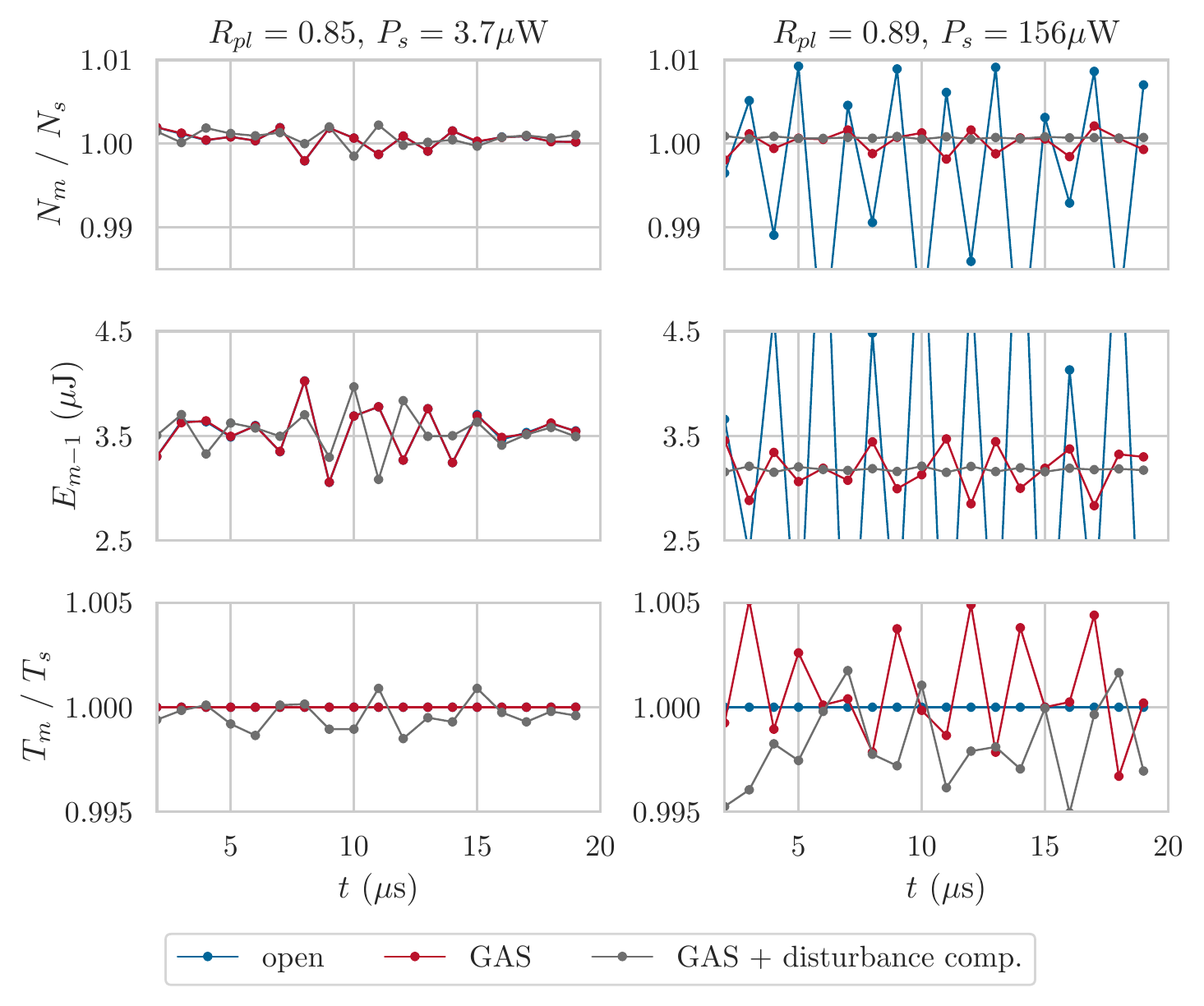} 
\caption{Populations (top), pulse energies (center) and control inputs (bottom) obtained from running the control algorithms presented in this paper. When the prelasing level $R_{pl}$ is chosen larger, the open loop becomes unstable while the GAS controller with disturbance compensation achieves its goal.} 
\label{fig:controller_comparison}
\end{center}
\end{figure}

In the following, we test the disturbance estimator from Sec. \ref{section:Real-time_disturbance_filter} coupled with the discrete-time control law derived in Secs. \ref{section:nonlinear_gas_controller} and \ref{section:linear_disturbance_feedback} in a simulation study using \textsc{Matlab / Simulink}. In all simulations, the laser operates at a repetition rate $f_{switch}$ = 1 MHz, pump power $P_p $ = 22.75 W, a nominal high-Q phase $T_s$ = 200 ns and a prelasing window $T_{pl}$ = 500 ns. For the given parameters and $R_{pl}\geq 0.87$, the uncontrolled laser model (\ref{equ:CTNM_avg}) becomes dynamically unstable. Therefore, the primary focus will be on the synergy between the nonlinear stabilisation of the steady-state population $N_s$ and the estimation and compensation of the disturbance $P_m(N_m,T_m)$. Furthermore, we will investigate the limitations of semi-active prelasing since it is not a priori clear if or under which circumstances semi-active prelasing as described in Sec. \ref{section:pulse-to-pulse_dynamics} is suitable for the control task discussed in Sec. \ref{section:Control_task}.
We will first check the disturbance estimator because it was obtained by employing some approximations and builds the foundation of the superordinate discrete-time controller. In Fig. \ref{fig:Prelasing_Kalman_illustration}, we show simulation data $P(t)$ in red, the noisy measurement $y(t)$ in light blue and the estimator with quasi-stationary covariance $C_\infty(t)$ reconstructing $\hat{P}(t)$ from the measurements in blue. Despite substantial sensor noise, the signal is reconstructed well. The yellow lines around the blue line are 20 other runs of the disturbance estimator, and the variance of $\hat{P}(\bar{t}_m)$ is approximately equal to that of $P(\bar{t}_m)$ resulting from the noise $\Sigma(t)$ in (\ref{equ:CTNM_avg_P}). \\   
To assess the nonlinear feedback controller and the disturbance compensation from Sec. \ref{section:controller_and_estimator_design}, we first draw closed-loop dynamics curves $f(N_m,P_m,T_m)$ with $T_m = g(N_m) + k(N_m)\left( P_m - P_s(N_m,g(N_m)) \right)$ for different $P_m$ and check whether the stability condition (\ref{equ:stability_condition}) and the disturbance rejection condition (\ref{equ:demanded_P_dead_beat}) are fulfilled. Fig. \ref{fig:controller_design} shows the normalised control inputs and resulting open- and closed-loop dynamics curves for $P_m$ being above, equal to and below $P_s(N_m,g(N_m))$ as well as $N_m \in [0.9 N_s, 1.1 N_s]$. Firstly, we observe from the lower subfigure of Fig. \ref{fig:controller_design} that the control input, i.e. the high-Q phase realistically varies by up to about 5 percent, which corresponds to the high-Q phase being 10 nanoseconds shorter or longer at most compared to $T_s=500$ ns. Secondly, the open-loop dynamics represented by dashed lines in Fig. \ref{fig:controller_design} vary greatly among different $P_m$ and exhibit the typical sharply falling slopes in $N_m$ that prevent global asymptotic stability (see (\ref{equ:stability_condition})). The closed-loop dynamics (solid lines), on the other hand, globally maintains a positive stability margin as intended by (\ref{equ:desired_closed_loop_slope}). With the linear disturbance compensation from Sec. \ref{section:linear_disturbance_feedback}, the influence of variations in $P_m$ on the system is also entirely mitigated as the three closed-loop curves coincide. \\
Having confirmed that the control and estimation components designed in Sec. \ref{section:controller_and_estimator_design} each fulfil their designated tasks, we can run simulations of (\ref{equ:CTNM_avg}) together with (\ref{equ:P_out_CTNM}) and (\ref{equ:E_out}) to investigate the system's behaviour employing a static high-Q phase $T_s$ (open loop), the nonlinear GAS with disturbance feedback using estimates from the disturbance filter, and just the GAS controller without disturbance compensation. In Fig. \ref{fig:controller_comparison}, we present discrete-time populations $N_m$, the previous pulse's energies $E_{m-1}$ and control inputs $T_m$ using the three different approaches. Note that when the system is in the state $N_m$, one can only know $E_{m-1}$ as a switching cycle needs to be completed before the outgoing pulse can be measured. The two columns in Fig. \ref{fig:controller_comparison} differ by the value of $R_{pl}$ which was used. A difference of four percent increases the nominal $P_s := P_s(N_s,T_s)$ by a factor of about 50 and has far-reaching consequences. In the left column, the system is dynamically stable. Hence, the GAS-controlled run does not differ from the open-loop behaviour since the GAS controller was designed minimally invasive and has no information about $P_m$. Even though the disturbance-compensating controller does set control actions in an effort to mitigate stochastic variations, the estimate $\hat{P}_m$ is overshadowed by stochastic influences after $\bar{t}_m$ which indicates that $R_{pl}$ was chosen too low. On the right-hand side, $R_{pl}$ is chosen more appropriately such that the disturbance estimator delivers relevant stochastic information to the controller. Consequently, the combined control concept manages to stabilise the pulse-to-pulse dynamics and reject stochastic fluctuations. The GAS controller by itself achieves dynamic stability, but undergoes stochastic fluctuations in the pulse energies of about 15 percent, while the open-loop system is dynamically unstable and enters a complicated limit cycle. Note that $R_{pl}$ on the right side of Fig. \ref{fig:controller_comparison} was, for demonstration purposes, chosen too large such that the average pulse energy is notably decreased. 
While it is always advisable to stabilise lasers which are nominally unstable, this simulation example and especially a comparison of the grey closed-loop pulse energies in the center of Fig. \ref{fig:controller_comparison} suggests that a system driven into instability due to the choice of $R_{pl}$ and then stabilised using feedback control might show superior performance to an operating point which is stable from the beginning.

\section{Conclusions and outlook}
\label{section:conclusion_and_outlook} 

In this work, a model-based nonlinear feedback control concept was presented which, based on a semi-active prelasing approach, stabilises the pulse-to-pulse dynamics of actively Q-switched lasers and globally mitigates stochastic influences from amplified spontaneous emission. In a simulation study, it was revealed that the choice of the prelasing level $R_{pl}$ is critical for the feasibility of real-time disturbance estimation and compensation.
To deploy the presented control approach on a real-world laser device, it is essential to develop parameter estimators which identify parameters of (\ref{equ:CTNM_avg}) since in real laser systems, many parameters are not known accurately enough or exhibit temperature-induced drifts. Since $N(t)$ and hence $N_m$ are in many cases not directly measurable, a state observer must harness pulse energies or similar output quantities to estimate $N_m$. Finally, implementing all control and estimation algorithms for the desired repetition rates is a challenging task. Currently, suitable control hardware and implementations are being investigated.

\bibliography{ifacconf}             

\end{document}